\documentclass[aps,prl,twocolumn,a4paper,10pt]{revtex4-1}

\usepackage[T1]{fontenc}		
\usepackage[english]{babel}		
\usepackage[latin1]{inputenc}	
\usepackage{times}

\usepackage{amsmath}			
\usepackage{amssymb}			
\usepackage{bbm}				

\usepackage[colorlinks=true, a4paper=true, pdfstartview=FitV,linkcolor=blue, citecolor=blue, urlcolor=blue]{hyperref}

\usepackage{graphicx}			
\usepackage{graphics}			
\DeclareGraphicsExtensions{.pdf}
\usepackage{color}		

\newcommand{\bea}{\begin{eqnarray}}
\newcommand{\eea}{\end{eqnarray}}

\newcommand{\bk}{\mathbf{k}}
\newcommand{\bq}{\mathbf{q}}

\newcommand{\bb}{\mathbf{b}}

\newcommand{\ba}{\mathbf{a}}
\newcommand{\bx}{\mathbf{x}}

\begin{document}
\title{Symmetry Enforced Stability of Interacting Weyl and Dirac Semimetals}

\author{Johan~Carlstr\"om and Emil~J. Bergholtz}
\affiliation{Department of Physics, Stockholm University, 106 91 Stockholm, Sweden}
\date{\today}

\begin{abstract}
The nodal and effectively relativistic dispersion featuring in a range of novel materials including two-dimensional graphene and three-dimensional Dirac and Weyl semimetals has attracted enormous interest during the past decade. Here, by studying the structure and symmetry of the diagrammatic expansion, we show that these nodal touching points are in fact perturbatively stable to all orders with respect to generic two-body interactions. For effective low-energy theories relevant for single and multilayer graphene, type-I and type-II Weyl and Dirac semimetals as well as Weyl points with higher topological charge, this stability is shown to be a direct consequence of a spatial symmetry that anti-commutes with the effective Hamiltonian while leaving the interaction invariant. A more refined argument is applied to the honeycomb lattice model of graphene showing that its Dirac points are also perturbatively stable to all orders. We also give examples of nodal Hamiltonians that acquire a gap from interactions as a consequence of symmetries different from those of Weyl and Dirac materials.    
\end{abstract}
\maketitle

{\it Introduction.---}
Graphene, Weyl semimetals and other Dirac materials are semi-metallic systems with 
a band structure that is gapped everywhere except at a set of "nodal points" in the Brillouin zone, where the valence and conduction bands meet \cite{doi:10.1080/00018732.2014.927109,HOSUR2013857,2017arXiv170501111A,doi:10.1146/annurev-conmatphys-031016-025225,volovik2009universe}. This band structure can arise due to lattice symmetry, spin-orbit coupling and breaking of time-reversal/inversion symmetry, or a combination of these, and results in low energy excitations that are mass-less quasi particles with an effective relativistic dispersion.  

Owing to their unique band structure, Dirac materials exhibit a number of novel electronic properties. 
Some of these were considered long before the discussion of Dirac/Weyl fermions in the context of condensed matter systems, notably the chiral anomaly, which is a magnetic response where Weyl nodes act as sources and sinks of a spontaneous current \cite{HOSUR2013857,PhysRev.177.2426, 2017arXiv170501111A}. A much more recent finding that is currently not well understood theoretically is Titanic magnetoresistance (the largest recorded so far) in the type-II Weyl semimetal WTe$_2$ \cite{10.1038/nature15768,10.1038/nature13763}.
Other topics that have attracted broad attention in Dirac systems include Fermi arcs and exotic surface states \cite{2017arXiv170501111A,doi:10.1146/annurev-conmatphys-031016-025225}. 

Thus, the wide interest in Dirac materials can be traced both to exotic and potentially useful electronic properties and emergent physics, but perhaps also to the fact that these phenomena appear already at the level of free fermions while interactions only seem to have a quantitative effect in many cases 
\cite{RevModPhys.84.1067,PhysRevB.79.085116,PhysRevB.96.195157}. 
This is notable for a number of reasons. First of all, many Dirac systems have touching points that are not topologically protected, meaning that an infinitesimal correction to the dispersion could open up a gap, leading to an insulating state. It is only in the three-dimensional Weyl semimetals that these points are actually protected \cite{doi:10.1146/annurev-conmatphys-031016-025225}. 
Secondly, interactions can in some cases be quite strong. In graphene, quantum Monte Carlo simulations suggest that the short range part of the Coulomb repulsion puts the system close to the Mott-insulating state \cite{PhysRevLett.111.056801,PhysRevB.89.195429}.
Subsequent work based on Diagrammatic Monte Carlo concludes that strong long-range interactions renormalise the Fermi velocities, but that the Dirac liquid remains stable \cite{PhysRevLett.118.026403}. 

In this work we examine the premises for stability of semi-metallic systems in the presence of inter-particle interactions, and show that these can be related to symmetries of the dispersion relation and the structure of the diagrammatic expansion.  Specifically, it is possible to define classes of theories where all diagrammatic corrections to the density vanishes at the touching points, thus preventing a gap from opening at all orders of perturbation theory.

{\it Diagrammatic treatment.---}
The Greens function of an interacting system can be written as a perturbative expansion in the interaction part of the Hamiltonian:
\bea\nonumber
G(\tau_a-\tau_b, \mathbf x_a-\mathbf x_b)=
Z^{-1}\sum_n \frac{(-1)^n}{n!} \int_0^\beta d \tau_i \; 
\\
\text{Tr}\{e^{-\beta H_0 } T[H_1(\tau_1)...H_1(\tau_n) \Psi^\dagger(\tau_a,\mathbf x_a)\Psi(\tau_b,\mathbf x_b)] \}\label{expansion}
\eea
which can be expressed as a set of connected diagrams \cite{fetter}. 
We can define a class of (dressed) tadpole insertions that take the form 
\bea
V_{\alpha\beta}(k=0) G_{\alpha\alpha}(\tau\to 0^-,k=0) \Psi^\dagger_\beta(\tau,\mathbf x) \Psi_\beta(\tau,\mathbf x).
\eea
The effect of these terms is merely to shift the chemical potential, meaning that if we are interested in the system at a specific stoichiometry these can be dropped. 
We thus obtain an interaction Hamiltonian of the form
\bea
H_1'=H_1-H_{\text{tadpole}}.
\eea
From the point of view of a diagrammatic treatment, this is equivalent to rejecting contributions from topologies where a boson is emitted without being reabsorbed, and this in turn rules out any propagators that close on themselves. 

From Eq. (\ref{expansion}) we make the crucial observation that, assuming two-body interactions, all terms in the self-energy have an odd number of propagators. We can thus write the contribution of a given diagram topology and set of internal variables as 
\bea
\delta \Sigma_{\alpha,\beta}(\omega,\mathbf k) =  \Pi_{i=1}^{2N-1} G^0_{\alpha_i\beta_i}(\omega_i,\mathbf k_i) \Pi_{j=1}^{N} V(\mathbf k_j)   ,\label{dSigma}
\eea
where $N$ is the expansion order. 

{\it Elementary argument for the effective low energy theory.---}
For low-energy descriptions of the band structure that are essentially expansions in $\bk$ around a touching point, we find a number of effective models for which there exists a mapping $T_-$ of $\mathbf k$  such that 
\bea
H_0(T_- \mathbf  k)=-H_0(\mathbf  k),\;\;\; |T_- \mathbf  k|=|\mathbf k|.\label{Tminus}
\eea
An example of this is a single Weyl cone, which is in the condensed matter context generally both anisotropic and tilted, and thus described by \cite{PhysRevB.91.115135}
\bea
H_0(\bk) =\sum_{ij} k_i v_{ij} \sigma_j +\sum_i k_ia_i \sigma_0,
\eea 
where $T_-$ corresponds to inversion, $\bk \to\ -\bk$.
Generalising this to the case of higher topological charge allows nodes on the form \cite{PhysRevLett.108.266802}
\bea
H_0(\bk) = k_z \sigma_z + (k_x+ik_y)^n\sigma_+ + (k_x-ik_y)^n\sigma_-,
\eea
where $T_-$ corresponds to a combination of rotation around the $z-$axis and reflection in the $xy-$plane.
Likewise, in graphene we obtain nodes on the form
\bea
H_0(\bk) = (k_x+ik_y)^n\sigma_+ + (k_x-ik_y)^n\sigma_-,
\eea
where $n=1$ for single layer and $n=2$ for bilayer systems \cite{0034-4885-76-5-056503}. Here $T_-$ corresponds to rotation around the $z$-axis by an angle $\phi=\pi/n$. 
Similar examples arise in 3D Dirac systems \cite{PhysRevB.84.235126}.

The presence of a $T_-$ symmetry leads to the nodal points being protected from interactions. This follows by applying (\ref{Tminus}) to the non-interacting Greens function:
\bea
G^0(\omega, T_- \mathbf k)=\frac{1}{i\omega -H_0(T_- \mathbf k)}=-G^0(-\omega , \mathbf k).\label{OddG0}
\eea
Inserting this into (\ref{dSigma}) gives
\bea
\delta \Sigma(-\omega,T_- \mathbf k)=-\delta \Sigma(\omega, \mathbf k)\label{dS_o}\ ,
\eea
where the minus sign results from the fact that we have an odd number of propagators in all diagram topologies. 
The Greens function of the interacting system is given by
\bea
G(\omega,\bk)=\frac{1}{i\omega -H_0(\bk)-\Sigma(\omega,\bk)}.\label{fullG}
\eea
At the nodal point the dispersion vanishes since $H_0(\bk=0)=0$, according to (\ref{Tminus}). From (\ref{dS_o}) we obtain that $\Sigma(\omega=0,\bk=0)=0$, and so $G(\omega,\bk=0)$ has a pole in $\omega=0$, which corresponds to a quasi particle at the Fermi surface, implying that the system remains gapless. The correction to the density at $\bk=0$ is given by 
\bea
\Delta\rho(\bk=0)=\sum_\omega \frac{1}{i\omega -\Sigma(\omega,\bk=0)}-\frac{1}{i\omega},
\eea 
which vanishes since the term in the summation is an odd function of $\omega$.
Thus, we conclude that given the existence of a mapping of the form (\ref{Tminus}), the nodal point is symmetry protected and the system remains semi-metallic.

{\it Refined argument for lattice models.---}
It is possible to generalise the above treatment to include other symmetries than (\ref{Tminus}) in order to address more complicated models. As an example we may consider graphene. In this material, the carbon atoms are arranged in a honeycomb lattice that can be divided into two triangular sub-lattices; $A$ and $B$. 
The Hamiltonian takes the form $H=H_0+H_1$, where
\bea
H_0=-t \sum_{\langle i,j \rangle,\sigma} (a_{i,\sigma}^\dagger a_{j,\sigma}+cc) +\sum_{i,\sigma} \mu n_{i,\sigma},\\
H_1=\frac{1}{2} \sum_{i,j,\sigma,\sigma'} V_{\sigma,\sigma'}(|\mathbf x_i-\mathbf x _j|)n_{i,\sigma}n_{j,\sigma'}.
\eea
When the fermi level is at the nodal point (i.e. $\mu=0$), we can parameterise this in terms of a pseudo-spin associated with the two sub-lattices according to

\[
H_0(\mathbf k)
=
\begin{bmatrix}
    0 & \Delta(\mathbf k)\sigma_0  \\
    \Delta^*(\mathbf k)\sigma_0 & 0
\end{bmatrix}
,\;\;\;\Delta(k)= \sum_{i=1}^3 e^{i \mathbf k\cdot\mathbf  \delta_i}
\]
where $\{\mathbf \delta_i\}_{i=1,2,3}$ are nearest neighbour vectors of the $A$-lattice, $\sigma_0$ is the $2\times2$ identity matrix acting on real spin, and the blocks correspond to pseudo-spin associated with different sub-lattices \cite{RevModPhys.81.109}. 
The bilinear part of the Hamiltonian has zeros in $\mathbf k_0=\pm (2\mathbf b_1-\mathbf b_2)/3$, where the $\mathbf b_i$ are the reciprocal lattice vectors. 

We can shift a Dirac point to the origin by the following change of variables:
\bea
\mathbf k=\tilde{\mathbf k}+\mathbf k_0\to\;\;\; \Delta(\tilde{\mathbf k}=0)=0.	\label{translate}
\eea
Rotating $\tilde{\mathbf k}$ by an angle $\frac{2\pi}{3}$ in the $xy$-plane we find
\bea
\Delta(R_{\frac{2\pi}{3}} \tilde{\mathbf k})=\Delta( \tilde{\mathbf k})e^{\pm i\frac{2\pi}{3}},
\eea
where the sign of the phase shift depends on the chirality of the Dirac point in $\mathbf k=\mathbf k_0$ (see supplementary material). This in turn implies that 
\bea
H_0(R_{\frac{2\pi}{3}} \tilde{\mathbf k})= e^{\pm \frac{i}{2} \frac{2\pi}{3}\tau_z}    H_0(\tilde{\mathbf k})   e^{\mp \frac{i}{2} \frac{2\pi}{3}\tau_z}\label{Rfermi},
\eea
so that a discrete rotation in $k$-space around the Dirac point can be related to a corresponding rotation of the pseudo-spin basis. 
This results in the following symmetry of the Greens function:
\bea\nonumber
G_0(i\omega,R_{\frac{2\pi}{3}}\tilde{\mathbf k})=
\frac{1}{i\omega-H_0(R_{\frac{2\pi}{3}}\tilde{\mathbf k})}\;\;\;\;\;\;
\\
=\frac{i\omega +H_0(R_{\frac{2\pi}{3}}\tilde{\mathbf k})}{-\omega^2-H^2_0(R_{\frac{2\pi}{3}}\tilde{\mathbf k}))}
= \frac{i\omega +e^{\pm \frac{i \pi}{3} \tau_z}H_0(\tilde{\mathbf k}) e^{\mp \frac{i\pi}{3}\tau_z}}{-\omega^2-H^2_0(\bk)}, \;\;\;\;\;\;\label{RG0}
\eea
where we have used the fact that $H_0^2(\bk)$ is diagonal, and thus commutes with the spin rotation. Since $i\omega$ commutes with $\tau_z$, discrete rotations around a Dirac point in $\bk-$space translates to a rotation of the pseudo-spin basis of the noninteracting Greens function as well. 

\begin{figure}[!htb]
\includegraphics[width=\linewidth]{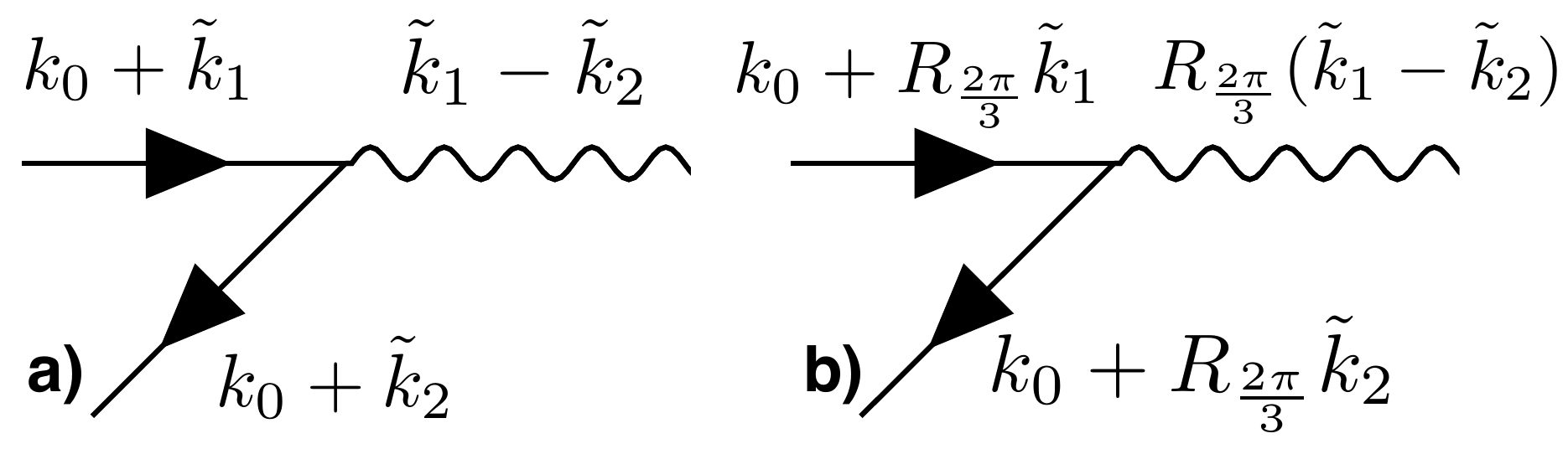}
\caption{
Fermionic and bosonic lines under translation and rotation: Shifting the momentum of fermions according to (\ref{translate}), the bosons are invariant as they are independent of $\mathbf k_0$ (a). Consequently, rotating all fermions around $\mathbf k=\mathbf k_0$, the bosons rotate around $\mathbf k=0$ (b).    
}
\label{VertexRotation}
\end{figure}

Conducting a diagrammatic expansion for the self-energy of the interacting theory we once again obtain corrections of the form (\ref{dSigma}). Rotating these around $\bk=\bk_0$, the fermionic lines transform according to (\ref{RG0}). By contrast, the bosonic (i.e. interaction) lines do in this case not rotate around the Dirac point, but rather around the origin. This follows from conservation of momentum at the vertices as illustrated in Fig. (\ref{VertexRotation}).
Thus, the rotated corrections take the form

\bea\nonumber
\delta \Sigma_{\alpha\beta}(\omega,R_{\frac{2\pi}{3}}\tilde{\mathbf k})=\Pi_{j=1}^{N} V(R_{\frac{2\pi}{3}}\bk_j)
\Pi_{i=1}^{2N-1}\\
\Big[
\frac{i\omega_i +e^{\pm \frac{i\pi}{3}\tau_z}H_0(\tilde{\mathbf k}_i)e^{\mp \frac{i\pi}{3}\tau_z}}{-\omega^2-H_0^2(\tilde{\bk}_i)}\Big]_{\alpha_i\beta_i}.
\eea
Due to lattice symmetry, the bosonic lines are invariant under the rotation (see supplementary material). The fermionic product can then be decomposed in diagonal frequency dependent factors and off-diagonal frequency independent factors according to
\bea\nonumber
\Pi_{i=1}^{2N-1} \big[i\omega_i +e^{\pm \frac{i\pi}{3} \tau_z} H_0(\tilde{\mathbf k}_i)e^{\mp \frac{i\pi}{3}\tau_z}\big]_{\alpha_i\beta_i}\;\;\;\\\nonumber
=\Pi_{a=1}^{N_a}i\omega_a  
\Pi_{b=1}^{N_b} \big[e^{\pm  \frac{i\pi }{3}\tau_z} H_0(\tilde{\mathbf k}_b)e^{\mp  \frac{i\pi}{3}\tau_z}\big]_{\uparrow\downarrow}
\;\;\;\\\nonumber
\times
\Pi_{c=1}^{N_c} \big[e^{\pm  \frac{i\pi }{3}\tau_z} H_0(\tilde{\mathbf k}_c)e^{\mp  \frac{i\pi}{3}\tau_z}\big]_{\downarrow\uparrow}
\;\;\;\\
=\Pi_{a=1}^{N_a}i\omega_a  
\Pi_{b=1}^{N_b} H^0_{\uparrow\downarrow}(\tilde{\mathbf k}_b)
\Pi_{c=1}^{N_c}  H^0_{\downarrow\uparrow}(\tilde{\mathbf k}_c)]
e^{\pm  \frac{i\pi }{3}(N_b-N_c)} \;\;\;\label{phaseFactor}
\eea
where ($\uparrow,\downarrow$) refer to the pseudo-spin degree of freedom associated with the two sublattices. 
Now, we can relate $\{N_i\}$ to different matrix elements of the self-energy as follows:

If $N_b-N_c=0$, then, the incoming and outgoing fermions have the same spin, meaning that the correction is to a diagonal element of the self-energy, i.e. of the form $\delta \Sigma_{\alpha\alpha}$, and is odd in frequency.   
 
If $N_b-N_c=\pm1$, then the correction is of the form $\delta \Sigma_{\uparrow\downarrow}$ or $\delta \Sigma_{\downarrow\uparrow}$, and even in frequency. 

From this and (\ref{phaseFactor}) we can draw the following conclusions: First of all, the transformation properties of the self-energy under rotation are the same as for the noninteracting Hamiltonian, i.e.
\bea
\Sigma(i\omega, R_{\frac{2\pi}{3}}\tilde{\mathbf k}) =e^{\pm  \frac{i\pi }{3}\tau_z} \Sigma(i\omega, \tilde{\mathbf k}) e^{\mp  \frac{i\pi }{3}\tau_z}. \label{RG}
\eea
Secondly, the diagonal corrections to the self-energy are odd in frequency, while the off-diagonal terms are even. 

From (\ref{RG}) it follows that the off-diagonal terms in $\Sigma(i\omega, \tilde{\mathbf k}=0)$ are zero, and thus that the self-energy is diagonal, and odd in frequency, i.e.
\bea
\Sigma(\omega,\tilde{\bk}=0)=-\Sigma(-\omega,\tilde{\bk}=0). \label{oddSigma}
\eea 
As in (\ref{fullG}), this gives a full Greens function of the form 
\bea
G(\omega,\tilde{\bk})=\frac{1}{i\omega -H_0(\tilde{\bk})-\Sigma(\omega,\tilde{\bk})}
\eea
which at the nodal point $\tilde{\bk}=0$ has a pole at zero energy and vanishing corrections to the density. 
Thus we find that the touching point is perturbatively stable to all orders.

{\it Unstable nodal points.---}
Breaking the $T_-$ symmetry, it becomes possible to either shift the nodal point, or to open up a gap. 
Conducting a first order diagrammatic expansion and discarding the Hartree term (which is a tadpole diagram that simply shifts the chemical potential), the self-energy is given by the Fock term:
\bea\nonumber
\Sigma^{\text{Fock}}_{\alpha\beta}(\bk) =-\frac{1}{2}\int d\bq  V(\bq)\\\nonumber
\times\Big( G^0_{\alpha\beta}(\delta,\bk-\bq)+G^0_{\alpha\beta}(-\delta,\bk-\bq) \Big) \\
=\int d\bq V(\bq)\sum_{\omega}\frac{i\omega+H^0_{\alpha\beta}(\bk-\bq)}{(\omega^2+ H_0^2(\bk-\bq))}
\eea
where we note that $H_0^2(\bk)$ is proportional to the identity matrix if $H_0$ is a particle-hole symmetric $2\times 2$ Hamiltonian.   
Hence, we find in this case that the self-energy only contains Pauli matrices already present in $H_0$. We now insert a simple Hamiltonian with a nodal touching point that is not topologically protected, and violates $T_-$ symmetry:
\bea\nonumber
H_0=\bk^2\sigma_z \to \Sigma^{\text{Fock}}_{\alpha\beta}(\bk) 
=-\frac{\delta_{\alpha\beta}}{2}\\\nonumber
\times\int d\bq V(\bq) \Big(\frac{1}{1+e^{\epsilon_\alpha(\bk-\bq)}}-\frac{1}{1+e^{-\epsilon_\alpha(\bk-\bq)}}\Big)\\
=\frac{\delta_{\alpha\beta}}{2}\int d\bq V(\bq)\tanh\Big(\frac{\epsilon_\alpha(\bk-\bq)}{2}\Big)
\eea 
where $\epsilon_{\alpha}(\bk)=\pm \bk^2$ (with the sign depending on $\alpha$). 
With $V(\bk)\ge 0$, the dispersions are shifted in opposite direction, away from the Fermi surface, thus opening a gap. This result can be readily generalised. With terms of the form $f(k_i)\sigma_i$, where $f(k_i)$ is even, gaps can open already at the level of first order perturbation theory. However if $f(k_i)$ breaks $T_-$ symmetry but changes sign, then the Fock diagram does not necessarily open a gap, but can instead merely change the position of the nodal point in $\bk$-space, and also relative to the Fermi surface.

{\it Discussion.---}
In conclusion, we have provided a general diagrammatic framework for studying the interplay between electronic interactions and symmetry in nodal semimetals.
For a wide class of Dirac materials we find that at the level of perturbation theory, the touching point is protected by the symmetry of the dispersion as manifested in the structure of the diagrammatic expansion to all orders. Thus there is no way to perturbatively open up a gap as a result interactions. 

This result has a number of implications for Dirac materials. 
First of all, it can explain why these systems are stable in the presence of interactions, despite the fact that the nodal points are in many cases not protected by topology and could be gaped by infndinitsmal perturbations. This is especially striking given the fact that these systems are frequently addressed in the non-interacting picture.
Second, it provides a simple and immediately identifiable condition under which a theory can be determined to be perturbatively stable. This argument can be readily generalised to more complicated models as in the case with the graphene lattice model.
Third, this work provides some insight into the question of diagrammatic treatment of Dirac materials in general. A natural expectation is that semi-metallic systems are ideal candidates for these techniques because of their vanishing density of states at the Fermi level.
Here we find further support for this scenario, namely that in the small region where the bands are close to the Fermi level, corrections fall of and even disappear at the touching point (while being practically irrelevant far away from the Fermi surface). 
This is in agreement with numerical work based on diagrammatic simulation techniques \cite{PhysRevLett.118.026403} that work extremely well for graphene. Interestingly, it has also been shown that for sufficiently weak short range interactions, the perturbative expansion for graphene is analytic down to zero temperature \cite{PhysRevB.79.201403}.

It should be stressed however that perturbative stability, even to infinite order, does not rule out a gap opening due to non-perturbative effects. This includes for example Mott instability in lattice models as well as pairing or charge density waves, which might be especially important in over-tilted type-II Weyl semimetals with a finite density of states at the Fermi level \cite{10.1038/nature15768,PhysRevB.96.201101}. In multilayer graphene nodes may split due to spontaneous symmetry breaking, though in this case the system remains semi-metallic \cite{PhysRevB.81.041401}.  

Another way of opening a gap is through explicit symmetry breaking. Even if the noninteracting part of the Hamiltonian satisfies (\ref{Tminus}), it is possible to introduce interaction terms that are not invariant under 
$\bk \to  T_- \bk $, so that (\ref{dS_o}) no longer holds.
In this context the three-dimensional Weyl semimetals stand out in the sense that the touching points are still stable due to the generic nature of the touching of two otherwise non-degenerate bands in three dimensions \cite{PhysRev.52.365}, while Dirac points in four-band models as well as two-dimensional Weyl points are generically unstable towards symmetry breaking owing to the accidental nature of the band-crossing. 
In most scenarios, interactions do however not explicitly break symmetries, and so the results reported here are central to understanding why the free-fermion picture can successfully describe a number of interacting quantum many-body systems. 

Our results also reveal a striking difference between the stability of nodal points versus interactions, as compared to the previously more extensively studied case of disorder \cite{PhysRevB.33.3263}. 
While we have shown here that symmetry is at the heart of the perturbative stability against interactions, the dimensionality and power law dispersion are the key factors when disorder is present. 
In particular, for a nodal point with a quasiparticle dispersion $\sim k^\alpha$ our work indicates stability 
whenever $\alpha$ is odd, while any even $\alpha$ will result in a gap. For a short-range disorder potential the corresponding statement is that---up to possible rare region effects \cite{PhysRevX.6.021042}---semi-metallicity is stable if $d > 2\alpha$ in $d-$dimensions, while the system becomes a diffusive metal for $d > 2\alpha$ \cite{PhysRevB.91.035133}.

Finally, our work raises intriguing questions about the free-fermion paradigm in Dirac systems. 
In particular, the equation (\ref{dS_o}) still allows imaginary terms in the self-energy that that are odd in frequency, and result in a finite quasi-particle lifetime. A natural questions is whether there are conditions under which these also vanish at the nodal point, thus giving rise to truly free electrons. 
Another interesting consideration is how the corrections to the density in the proximity of the touching point can be related to the noninteracting theory.

{\it Acknowledgments.---}
This work was supported by the Swedish research council (VR) and the Wallenberg Academy Fellows program of the Knut and Alice Wallenberg Foundation. 
The authors would like to thank Vladimir Juricic, Nikolay Prokof'ev and Boris Svistunov for important input and discussions. 

\bibliography{biblio}

\subsection{Supplementary material}

In the supplementary material we collect derivations of the rotational symmetry that do not fit in to the main paper. The first calculation concerns $H_0(\bk)$, and demonstrates that a discrete rotation by $2\pi/3$ around a nodal point in $\bk$-space can be related to a rotation of the pseudo-spin basis. 
The second derivation shows that a bosonic line is invariant under rotation of $2\pi/3$ around the origin. 

\subsection{Rotational symmetry of $H_0$}
In this section we demonstrate that discrete rotations around the nodal point in $\bk$-space are related to a rotations of the pseudo-spin basis according to:
\bea
H_0(R_{\frac{2\pi}{3}} \tilde{\bk})=e^{\pm\frac{i\pi }{3}\tau_z} H_0(\tilde{\bk}) e^{\mp\frac{i\pi }{3}\tau_z}
\eea
where $R_{\frac{2\pi}{3}}$ is a rotation by $\frac{2\pi}{3}$ in the $xy$-plane, $\tilde{\bk}$ is the distance from a nodal point in $\bk$-space and $\tau_z$ acts on the pseudo-spin.  
\\
\\
{\it Proof.---}
Let $\{\delta_i\}_{i=1,2,3}$ denote the nearest neighbour vectors of the $A$-lattice, and define $R_{\frac{2\pi}{3}}$ be a rotation in the $xy$-plane by $2\pi/3$. Then, the nearest neighbour vectors are related according to
\bea\nonumber
R_{\frac{2\pi}{3}} \delta_i=\delta_{i+1},\; i<3\\
R_{\frac{2\pi}{3}} \delta_3=\delta_{1},\; \label{delta_symmetry_sup}
\eea
The lattice vectors are then given by
\bea
\ba_1=\delta_1-\delta_2,\;\;\;
\ba_2=\delta_1-\delta_3.		\label{a_i_sup}
\eea
The rotational properties can be deduced from (\ref{delta_symmetry_sup}).
\bea\nonumber
R_{\frac{2\pi}{3}} \ba_1= \delta_2-\delta_3=\ba_2-\ba_1\\
R_{\frac{2\pi}{3}} \ba_2= \delta_2-\delta_1=-\ba_1\label{rot_ai_sup}
\eea

Defining $R_{\frac{\pi}{2}}$ to be a rotation in the $xy$-plane of $\pi/2$ we obtain the following reciprocal lattice vectors:
\bea\nonumber
\bb_1=2\pi\frac{R_{\frac{\pi}{2}} \ba_2}{\ba_1^\dagger R_{\frac{\pi}{2}} \ba_2},\;\;\;
\bb_2=2\pi\frac{R_{\frac{\pi}{2}} \ba_1}{\ba_2^\dagger R_{\frac{\pi}{2}} \ba_1} \label{b_i_sup}
\eea
satisfying 
\bea
\bb_i\cdot \ba_j=\delta_{ij}2\pi.\label{reciprocal_sup}
\eea
Their rotational symmetry can be obtained from (\ref{rot_ai_sup}) and noting that $\ba_1^\dagger R_{\frac{\pi}{2}} \ba_2=-\ba_2^\dagger R_{\frac{\pi}{2}} \ba_1$:
\bea\nonumber
R_{\frac{2\pi}{3}} \bb_1=2\pi\frac{R_{\frac{\pi}{2}} (-\ba_1)}{\ba_1^\dagger R_{\frac{\pi}{2}} \ba_2}=\bb_2,\\
R_{\frac{2\pi}{3}} \bb_2=2\pi\frac{R_{\frac{\pi}{2}} (\ba_2-\ba_1)}{\ba_2^\dagger R_{\frac{\pi}{2}} \ba_1}=-(\bb_1+\bb_2),\label{rot_bi_sup}
\eea

The tight-binding model can be written as 
\[
H_0(\bk)
=
\begin{bmatrix}
    0 & \Delta(\bk)\sigma_0  \\
    \Delta^*(\bk)\sigma_0 & 0
\end{bmatrix}
,\;\;\;\Delta(\bk)= \sum_{i=1}^3 e^{i \bk\cdot \delta_i}
\]
where $\{\delta_i\}_{i=1,2,3}$ are nearest neighbour vectors of the $A$-lattice, $\sigma_0$ is the $2\times2$ identity matrix acting on real spin, and 
the blocks correspond to pseudo-spin associated with different sub-lattices. 

The function $\Delta(\bk)$ has zeroes in $\bk'=(2\bb_1+\bb_2)/3$. This follows from the following considerations:

From (\ref{a_i_sup}) we find that $\delta_2=\delta_1-\ba_1$. Using (\ref{reciprocal_sup}) we then obtain 
\bea
\bk_0\cdot \delta_2=\bk_0\cdot \delta_1-\frac{4\pi}{3}.
\eea
Likewise, we get $\delta_3=\delta_1-\ba_2$ implying 
\bea
\bk_0\cdot \delta_3=\bk_0\cdot \delta_1-\frac{2\pi}{3},
\eea
resulting in cancellation so that $\Delta(\bk=\bk')=0$. Also, we have that $\Delta(\bk=-\bk')=\Delta^*(\bk=\bk')=0$, so that we have zeroes $\Delta(\bk=\bk_0),\;\bk_0=\pm (2\bb_1+\bb_2)/3 $.

We can introduce a translation in momentum space according to
\bea
\bk=\tilde{\bk}+\bk_0 \label{translation_sup}
\eea
so that $\Delta(\tilde{\bk}=0)=0$. We now proceed to see how $\Delta$ behaves under a discrete rotation around the Dirac point by $-\frac{2\pi}{3}$ denoted by $R^\dagger_{\frac{2\pi}{3}}$. For concreteness we consider the case $\bk_0=(2\bb_1+\bb_2)/3$. We obtain
\bea
\Delta(R^\dagger_{\frac{2\pi}{3}} \tilde{\bk})=\sum_i e^{i (R^\dagger_{\frac{2\pi}{3}}\tilde{\bk}+\bk_0)\cdot \delta_i}
\\
=\sum_i e^{i (\tilde{\bk}+R_{\frac{2\pi}{3}}\bk_0)\cdot R_{\frac{2\pi}{3}}\delta_i}
=\sum_i e^{i (\tilde{\bk}+R_{\frac{2\pi}{3}}\bk_0)\cdot \delta_i},
\eea
where in the last stage we have used (\ref{delta_symmetry_sup}). The rotational properties of $\bk_0$ can be deduced from the relations (\ref{rot_bi_sup}). We find 

\bea
R_{\frac{2\pi}{3}}\bk_0=R_{\frac{2\pi}{3}}\frac{2\bb_1+\bb_2}{3}=\frac{\bb_2-\bb_1}{3}=\bk_0-\bb_1,
\eea
implying
\bea
\Delta(R^\dagger_{\frac{2\pi}{3}} \tilde{\bk})=\sum_i e^{i (\tilde{\bk}+\bk_0-\bb_1)\cdot \delta_i}\label{rotDelta_sup}.
\eea
To evaluate $\bb_1\cdot \delta_i$ we first compute $\delta_1\cdot \bb_1$ explicitly using (\ref{delta_symmetry_sup}), (\ref{a_i_sup}) and (\ref{b_i_sup}):
\bea\nonumber
 \delta_1\cdot \bb_1=\delta_1\cdot 2\pi\frac{R_{\frac{\pi}{2}} (\delta_1-\delta_3)}{(\delta_1-\delta_2)R_{\frac{\pi}{2}} (\delta_1-\delta_3)}\;\;\;\;\;\;\\\nonumber
 =  2\pi\frac{-\delta_1\cdot R^2_{\frac{2\pi}{3}}R_{\frac{\pi}{2}}  \delta_1}{(\delta_1-R_{\frac{2\pi}{3}}\delta_1)R_{\frac{\pi}{2}} (\delta_1-R^2_{\frac{2\pi}{3}}\delta_1)}=\;\;\;\;\;\;\\
   \frac{-2\pi\cos(\frac{11\pi}{6})\delta_1^2} 
 {
 \delta_1^2\Big(\cos(\frac{\pi}{2})+\cos(\frac{7\pi}{6})-\cos(\frac{11\pi}{6})-\cos(\frac{\pi}{6})\Big)}\label{d1_b1_sup}
 =\frac{2\pi}{3}.\;\;\;\;\;\;
\eea
Next, we use (\ref{a_i_sup}) and (\ref{reciprocal_sup}) to obtain
\bea
(\delta_1-\delta_2)\cdot \bb_1=2\pi,\;\;\;
(\delta_1-\delta_3)\cdot \bb_1=0.
\eea
Together with (\ref{d1_b1_sup}) this gives 
\bea
\delta_i\cdot \bb_1=\frac{2\pi}{3}+n2\pi\label{di_b1_sup}.
\eea
Inserting (\ref{di_b1_sup}) in (\ref{rotDelta_sup}) we find 
\bea
\Delta(R^\dagger_{\frac{2\pi}{3}} \tilde{\bk})e^{i\frac{2\pi}{3}}=\Delta(\tilde{\bk})\to \Delta(R_{\frac{2\pi}{3}} \tilde{\bk})=\Delta(\tilde{\bk})e^{i\frac{2\pi}{3}}. \label{rotoD_sup}
\eea
This in turn implies that 
\bea
H_0(R_{\frac{2\pi}{3}} \tilde{\bk})=e^{\frac{i\pi }{3}\tau_z} H_0(\tilde{\bk}) e^{-\frac{i\pi }{3}\tau_z}
\eea
when the rotation is around $\bk_0=\frac{2\bb_1+\bb_2}{3}$. If the node has the opposite chirality, then the phase shift changes sign.

\subsection{Rotational symmetry of interaction lines}
Next we consider rotation of bosonic lines. When rotating the fermionic lines around a Dirac point, the bosonic lines rotate around the origin. This follows from momentum conservation, see Fig. (\ref{VertexRotation}).

Thus, we want to determine how $V(\mathbf{k})$ rather than $V(\tilde{\mathbf{k}})$ behaves under rotation, and this follows directly from symmetry:

\bea
V(\mathbf{k})=\sum_{i,j}V(|\bx_i-\bx_j|)e^{i(\bx_i-\bx_j)\cdot \bk}.
\eea
Under rotation this becomes 
\bea
V(R_{\frac{2\pi}{3}}\mathbf{k})=\sum_{i,j}V(|\bx_i-\bx_j|)e^{i(\bx_i-\bx_j)\cdot R_{\frac{2\pi}{3}}\bk}\\
=\sum_{i,j}V(|R^\dagger_{\frac{2\pi}{3}}(\bx_i-\bx_j)|)e^{iR^\dagger_{\frac{2\pi}{3}}(\bx_i-\bx_j)\cdot \bk}
\eea
but due to lattice symmetry, the sets $(\bx_i-\bx_j)$ and $R^\dagger_{\frac{2\pi}{3}}(\bx_i-\bx_j)$ are identical, and we conclude that
\bea
V(R_{\frac{2\pi}{3}}\mathbf{k})=V(\mathbf{k}),\label{rotV_sup}
\eea 
so that bosonic lines are invariant under a discrete rotation.

\end{document}